\documentclass[11pt]{article}

\usepackage{euscript}

\usepackage{latexsym}
\usepackage{amsmath, amssymb,graphics}
\usepackage[all]{xy}

\usepackage{fullpage}

\newtheorem{theorem}{Theorem}
\newtheorem{definition}[theorem]{Definition}

\newtheorem{proposition}[theorem]{Proposition}

\newtheorem{observation}{Observation}

\renewcommand{\(}{\begin{equation*}}
\renewcommand{\)}{\end{equation*}}
\newcommand{\bea}{\begin{eqnarray*}}
\newcommand{\eea}{\end{eqnarray*}}
\newcommand{\R}{{\mathbb R}}
\newcommand{\C}{{\mathbb C}}
\newcommand{\Z}{{\mathbb Z}}
\newcommand{\Q}{{\mathbb Q}}

\newcommand{\cF}{\ensuremath{\mathcal F}}

\newcommand{\bo}{\raise-1mm\hbox{\Large$\Box$}}              % D'Alembertian
\newcommand{\beq}{\begin{equation}}
\newcommand{\eeq}{\end{equation}}

\numberwithin{equation}{section}

\renewcommand{\(}{\begin{equation}}
\renewcommand{\)}{\end{equation}}

\newcommand{\RR}{{\mathbb R}}

\newcommand{\RP}{\RR \text{P}}

\def\R{{\mathbb R}}
\def\Z{{\mathbb Z}}
\def\Q{{\mathbb Q}}
\def\C{{\mathbb C}}

\def\1{{\bf 1}}

\def\<{\langle}
\def\>{\rangle}

\numberwithin{equation}{section}

\renewcommand{\(}{\begin{equation}}
\renewcommand{\)}{\end{equation}}
\begin{document}

\begin{titlepage}
%\begin{flushright}

%hep-th/xxxxxxx
%\end{flushright}

\vspace{2em}
\def\thefootnote{\fnsymbol{footnote}}

\begin{center}
{\Large\bf 
Twisted topological structures related to M-branes II: \newline
Twisted Wu and Wu${}^c$ 
structures}
\end{center}
\vspace{1em}

\begin{center}
Hisham Sati 
\footnote{e-mail: {\tt
hsati@pitt.edu} \newline
Research supported by NSF grant PHY }
\end{center}

\begin{center}
Department of Mathematics\\
University of Pittsburgh\\
Pittsburgh, PA 15260

\end{center}

\vspace{0em}
\begin{abstract}
\noindent
 
Studying the topological aspects of M-branes in M-theory leads to 
various structures related to Wu classes. First we interpret Wu classes
themselves as twisted classes and then define twisted notions of Wu 
structures. These generalize
many known structures, including Pin${}^-$ structures, 
twisted Spin structures in the sense of Distler-Freed-Moore, 
Wu-twisted differential cocycles appearing in the work of Belov-Moore,
as well as ones introduced by the author,
such as twisted Membrane and twisted String${}^c$ structures.
In addition, we introduce Wu${}^c$ structures, which generalize Pin${}^c$ structures,
as well as their twisted versions. We show how these 
structures generalize and encode the usual structures defined via 
Stiefel-Whitney classes.

\end{abstract}

\end{titlepage}

\tableofcontents

\section{Introduction}
%%%%%%%%%%%%%%%%%%%%%%%%%%%%%

M-theory has proven over the years to be a very rich sources of 
various mathematical structures. In previous work 
\cite{S1}  \cite{S3} \cite{S4}
we uncovered some geometric and topological 
structures related to M-branes in M-theory. 
This letter is a continuation of the work \cite{S4}, where
the emphasis is on twisted topological structures. 

\vspace{3mm}
The main focus will be on Wu classes $v_i$, which are mod 2 characteristic 
classes that can be written as polynomials over $\Z_2$ in the 
Steifel-Whitney classes $w_j$ of same and/or lower degrees.
In the lowest two degrees these structures are familiar: 
$v_1$ is the obstruction to manifold orientation, and 
$v_2$ is the obstruction to having a Pin${}^-$
structure. The latter structure has many interesting applications to 
low-dimensional topology \cite{KT}. 
In high dimensions, Wu classes have interesting applications 
to surgery problems.
In degrees four and six, the Wu classes have applications to the M5-brane 
partition function \cite{W-5} \cite{HS} and to topological aspects of 
type IIB string theory \cite{W-5} \cite{BeM}, respectively. 

\vspace{3mm}
What we do in this paper can be summarized as follows

\vspace{2mm}
\noindent {\bf 1.} We interpret Wu structures themselves already as twisted structures
defined by the Stiefel-Whitney classes. These structures include Spin 
structures and Membrane structures \cite{S4}.  This is done in section \ref{sec t}.

\vspace{2mm}
\noindent {\bf 2.} In section \ref{sec tw} we introduce the notion of {\it twisted Wu structure}, 
which in cohomological degree two is related  to twisted 
Spin structure (cf. \cite{Wa} \cite{DisFM}). In degree four this
will be a generalization of twisted Membrane structure, introduced
in \cite{S4}. 

\vspace{2mm}
\noindent {\bf 3.} 
We  also introduce structures that we call {\it Wu${}^c$ structures} in section \ref{sec w c}.
In the appropriate degrees, these are generalizations of the 
Spin${}^c$ structure and of the String${}^{K(\Z,3)}$ structure 
(cf. \cite{S4}).

\vspace{2mm}
\noindent {\bf 4.} Finally, in section \ref{sec tw c}, we describe a twist for the Wu${}^c$ structures
leading to {\it twisted Wu${}^c$ structures}. 
These are generalizations of twisted Spin${}^c$ (and twisted Pin${}^c$) structures 
and twisted String${}^{K(\Z,3)}$ structures \cite{S4} 
in degree 3 and degree 7, respectively. 

\vspace{2mm}
\noindent 
Throughout, we emphasize the motivation and the relation to M-branes via examples.
These example also include (spacetime) M-theory and type II string theory.

%%%%%%%%%%%%
\section{Wu classes and Wu structures}
%%%%%%%%%%%
We start by providing a basic description and properties 
of the Wu classes and Wu structures that will be used in later sections.

%%%%%%%%
\subsection{Wu classes}
%%%%%%%

\paragraph{The Stiefel-Whitney classes in terms of the Wu classes.}
Let $B$O denote the classifying space of the stable orthogonal group ${\rm O}=\bigcup_{k=0}^\infty
{\rm O}(k)$.
The $i$-dimensional {\it universal Wu class} $v_i$ is the element of 
$H^i(BO; \Z_2)$ defined inductively via the Steenrod square $Sq^j$ as 
(cf. \cite{Mi}\cite{SE} \cite{St1}) 
%\cite{ST}
\(
v_0=w_0=1 \quad {\rm and} \quad w_i= v_i + Sq^1 v_{i-1} + \cdots + Sq^i v_0~{\rm if}~ i \geq 1. 
\label{eq wu}
\)
These classes can also be defined for manifolds via the classifying map. 
Let $M^n$ be a closed $n$-dimensional manifold and 
consider the action of the Steenrod square on the cohomology of the manifold 
$Sq^i: H^n(M; \Z_2) \to H^{n+i}(M; \Z_2)$.
By Poincar\'e duality, there are unique classes $v_i \in H^i(M;\Z_2)$ satisfying 
\(
\langle v_i \cup x, [M] \rangle = \langle Sq^i x, [M] \rangle 
\)
for all $x \in H^{n-1}(M; \Z_2)$.
Thus the relation is (cf. \cite{MS} \cite{SY})
\(
v_0(M)=1 \quad {\rm and} \quad w_i(M)= v_i(M) + 
Sq^1 v_{i-1}(M) + \cdots + Sq^i v_0(M)~{\rm if}~ i \geq 1. 
\)
That is, if $f$ denotes the classifying map for the stable tangent bundle of $M$, then
$f^*w_i=w_i(M)$ and $f^*v_i= v_i(M)$  if $i\geq 0$.
 Since $Sq^i x=0$ for $i > n-i$, 
the classes vanish: $v_i=0$ for $i > [\frac{n}{2}]$. The class
\(
v=1 + v_2 + \cdots + v_{[\frac{n}{2}]} \in H^*(M; \Z_2)
\)
is  the {\it total Wu class} of $M$. 
The total Stiefel-Whitney class $w= 1 + w_1 + w_2 + \cdots + w_n$ of $M$
is determined by the {\it Wu formula} $w= Sq (v)$, where 
$Sq= 1 + Sq^1 + Sq^2 + \cdots$
is the total Steenrod square operation.
 The Wu formula can be expanded as
$
1 + w_1 + \cdots = Sq^0( 1 + v_1 + \cdots )
+ Sq^1(1 + v_1 + \cdots ) + \cdots$.
Via $w_i= \sum_{j=0}^i Sq^{i-j} v_j$, the first few classes are
\bea
w_1 &=& Sq^0(1) + Sq^0(v_1) = v_1\;,
\\
w_2&=& Sq^2(1) + Sq^1(v_1) + Sq^0(v_2)=v_1^2 + v_2\;,
\\
w_3&=& Sq^3(1) + Sq^2(v_1) + Sq^1(v_2) + Sq^0(v_3)= Sq^1(v_2) + v_3\;.
\eea

Since Stiefel-Whitney classes are more familiar than Wu classes, we
are generally more interested in inverting the above relations.

\paragraph{The Wu classes in terms of the Stiefel-Whitney classes.}
The above relation, i.e. the Wu formula, for the Steifel-Whitney classes in terms of the
Wu classes can be inverted to get the latter classes as polynomials 
in the former classes. We describe two ways of doing this. 
The first is to use the anti-automorphism (canonical conjugation) 
$\chi ({\cal A}(2))$ on the mod 2 Steenrod
algebra ${\cal A}(2)$. The second way uses the Todd classes Td${}_i$. 

\vspace{2mm}
\noindent {\bf 1.} {\it Anti-automorphism:}
One can invert the relation and write the total Wu class in terms of the total
Stiefel-Whitney class
\(
v(M)= \chi (Sq) w(M)\;.
\label{eq vw}
\)
The anti-automorphism $\chi$ is defined recursively using Thom's recursion 
formula
$
\sum_{i=0}^n Sq^i \chi (Sq^{n-i})=0$.
Since the mod 2 Steenrod algebra is generated multiplicatively 
by the elements $Sq^{2^n}$, $\chi$ is determined completely
by knowledge of $\chi (Sq^{2^n})$ for all $n$. The identity \cite{Str}
$
Sq^{2^n} + \chi (Sq^{2^n})= Sq^{2^{n-1}} \chi (Sq^{2^{n-1}})
$
and the formula 
$
\chi (Sq^{2^n})= Sq^{2^n} + \sum_{i=1}^{n-1} \prod_{i=1}^{n-1} 
\left(
\prod_{j=1}^i Sq^{2^{n-j}}
\right) Sq^{2^{n-i}}
$
allow for explicit calculation of $\chi$.
There is, in fact, an explicit formula for individual Wu classes in terms of the Stiefel-Whitney
classes \cite{Yo}. It follows from the Wu formula that 
$
v_n= \sum_{i=1}^n \theta^{n-i} w_i$,
where $\theta^l= \chi (Sq^l) \in {\cal A}(2)$ is the conjugation of 
$Sq^l$ in ${\cal A}(2)$ and is defined
inductively by
\(
\theta^l= Sq^l + \sum_{i=1}^{l-1} Sq^i \theta^{l-i}= 
Sq^l + \sum_{j=1}^{l-1} \theta^{l-j} Sq^j\;.
\) 
For $l=0$, $\theta^0=1$ and for $l=1$, $\theta^1=Sq^1$.  
The odd-dimensional terms are given in terms of the even-dimensional
ones via $\theta^{2n+1}= \theta^{2n} Sq^1$.

\vspace{2mm}
\noindent {\bf 2.} {\it Todd classes:}
The second method is much easier and uses the relation between
the Wu classes and the Todd classes, as polynomials in 
the Steiefel-Whitney classes
(and {\it not} in the Chern classes), i.e. 
${\rm Td}_n(w_1, \cdots, w_n)$. The formula is given by 
(see \cite{HBJ})
\(
v_n \equiv 2^n\cdot {\rm Td}_n(w_1, \cdots, w_n) \mod 2 \;,
\)
through which the expansion of the Wu classes can be most efficiently read off from the
corresponding expansion of the Todd genus. For example, for the low
degree classes of most direct relevance to us, we have 
\begin{eqnarray}
v_1&=& w_1\;,
\nonumber\\
v_2 &=& w_2 + w_1^2\;,
\nonumber\\
v_3&=& w_1 w_2\;,
\nonumber\\
v_4 &=& w_4 + w_3w_1 + w_2^2 + w_1^4\;,
\nonumber\\
v_5&=& w_4 w_1 + w_3w_1^2 + w_2^2 w_1 + w_2w_1^3\;,
\nonumber\\
v_6&=&  w_4 w_2  + w_4 w_1^2 +  w_3^2+ w_3 w_2 w_1 + w_3w_1^2 + w_2^2 w_1^2\;. 
\label{eq list}
\end{eqnarray}

\paragraph{Additivity property of Wu classes.} From relation 
\eqref{eq vw} , and using the 
formula $w(E \oplus F)= w(E) \cup w(F)$, we immediately see 
that 
$v(E \oplus F)= \chi (Sq) (w(E \oplus F)) = \chi (Sq) \left( w(E) \cup w(F) \right)$, 
which is equal to 
$\chi (Sq)(w(E)) \cup \chi (Sq)(w(F))$
so that the Wu class is additive under Whitney sum, i.e., 
it satisfies 
\(
v(E \oplus F)= v(E) \cup v(F)\;.
\)

\paragraph{Example 1.} Consider the second Wu class. 
We have $v_2(E \oplus F)$ equals to 
$w_2(E \oplus F) + w_1(E \oplus F) \cup w_1(E \oplus F)$. 
Expanding, the first summand gives $w_2(E) + w_2(F) + w_1(E) w_1(F)$,
while the second summand gives $w_1(E)^2 + w_1(F)^2 + 2 w_1(E) w_1(F)$.
Since $w$ is  a mod 2 class, we have $2w_1(E) w_1(F)=0$. Therefore, altogether
we have $w_2(E) + w_2(F) + w_1(E)^2 + w_1(F)^2 + w_1(E) w_1(F)$,
which can be written as $v_2(E) + v_2(F) + v_1(E) v_1(F)$.

 \vspace{3mm}
As we will consider structures defined by Wu classes, it would be useful to 
get some idea of when such classes vanish in relation to Stiefel-Whitney 
classes.

\paragraph{(Non)vanishing of the Wu classes.}
The Stiefel-Whitney numbers $w_{i_1} \cdots w_{i_r} [M^m]$, for $i_1 + \cdots i_r=m$,
 determine  the cobordism class of the $m$-manifold 
 $M$ \cite{St}. This implies, in particular, that if 
$n>0$ and $M$ is not a boundary then there must be an 
$i>0$ for which $v_i \neq 0$.  
Let $B_k^*=H^*(BO;\Z_2)/I(v_i~|~i >k)$, where $I(v_i~|~i>k)$ is the ideal 
generated over the Steenrod algebra by the classes $v_i$, $i>k$.
%, where $v=Sq^{-1} w$. 
For $k=1$, $B_1^*$ has $v=1+v_1$ and $w= Sq v= 1 + v_1 + v_1^2$ with 
$0=Sq^1(v_1^2)= Sq^1 w_2= w_3 + w_2 w_1= w_2 w_1=v_1^3$.
Thus $B_1^*=\Z_2 \oplus \Z_2 \oplus \Z_2$ with basis $1, v_1, v_1^2$.
It can be shown by induction that $B_k^*$ is finite-dimensional \cite{SY}.
Contrary to what happens with Stiefel-Whitney classes, 
there is a closed manifold $M^n$ which bounds with nonvanishing 
Wu class \cite{SY}
$v= 1 + v_1 + v_2 + v_3$ for $n>7$,
$v=1+ v_1 + v_2$ for $n> 5$,
and
$v= 1 + v_1$ for $n>2$.

\paragraph{Example 2.} 
Every closed seven-manifold $M^7$ which fibers over the sphere
$S^5$ or $S^6$ is a boundary. That is 
there are bundles 
$\Sigma_2 \buildrel{i}\over{\to} M^7 \buildrel{\pi}\over{\to} S^5$ for which
$v_i=0$ for $i>1$ and
bundles 
$S^1 \buildrel{i}\over{\to} M^7 \buildrel{\pi}\over{\to} S^6$ for which
$v_i=0$ for $i>0$.
Similarly, every closed eleven-manifolds $Y^{11}$ which fibers over the 
sphere $S^j$ with $7 \leq j \leq 10$ is a boundary. In this case
$v_i=0$ for $i>2$, $i>1$, $i>1$ and $i>0$, corresponding respectively
to the cases $j=7, 8, 9,$ and 10.

\paragraph{Odd-dimensional Wu classes.} 
The odd-dimensional Wu classes can be written in terms of the lower
Wu classes via the formula \cite{Yo}
\(
v_{2n+1}= \sum_{i \geq 1} (w_1)^{2^i -1} v_{2n+2-2^i}\;.
\)
This immediately implies that for oriented manifolds 
the odd-dimensional Wu classes are all zero. 
In fact, for $x \in H^{n-2i-1}(M)$, the same conclusion can be reached from 
the relation 
$
Sq^{2i+1}(x)= Sq^1 Sq^{2i} (x)
= v_1 \cup Sq^{2i}(x)
= w_1 \cup Sq^{2i}(x)
$.

\paragraph{Example 3. Special class of manifolds.} 
Consider manifolds $M$ whose total Stiefel-Whitney class
$w(M)$
has nonzero components only in degrees that are powers of 2, i.e.
it satisfies the condition 
$w(M)= 1 + \sum_{j\geq 1} w_{2^{j-1}}$. 
This is satisfied by a class of manifolds which include projective spaces. 
For example, for real projective spaces we have $w(\R P^5)=1+ a^2 + a^4$,
$w(\R P^9)=1 + a^2 + a^8$ and $w(\R P^{11} )= 1 + a^4 + a^8$, where 
$a$ is the class of the real classifying line bundle.
For such
manifolds, the Wu classes have
%\eqref{eq Yo} reduces to 
the explicit form \cite{Yo}
\bea
v_i&= & \sum_{j=1}^m (w_{2^{j-1}})^{2^{m-j}} \hspace{4.5cm} {\rm if~} i=2^{m-1} \geq 1\;,
\nonumber\\
v_i &=& \sum_{j=1}^m \sum_{k=m+1}^n (w_{2^{j-1}})^
{[(i-2^{k-1})/2^{j-1}]}
(w_{2^k})^{2^{k-j-1}} \quad {\rm if~} i=2^{m-1}+2^{n-1} {\rm~with~} n>m \geq 1\;,
\nonumber\\
v_i &=& 0 \hspace{6.7cm} {\rm otherwise}.
\eea
Indeed for the real projective space $\R P^n$, the total Wu class is 
$v(\R P^n)=\sum_{i=0}^n \binom{n-i}{i}a^i$ where $a$ is the class of the classifying
real line bundle. We will discuss the significance of powers of the Steifel-Whitney classes
(especially squares) in our context in section \ref{sec w c}.

%%%%%%%%
\subsection{Wu structures via classifying spaces}
%%%%%%%%
 \label{sec wu st}
Consider the principal fibration $BO[v_{2k}]$ over the classifying space
 BO of the orthogonal group, with fiber the Eilenberg-MacLane space 
 $K(\Z_2, 2k-1)$, and with Postnikov invariant of the fibration equal to the 
 Wu class $v_{2k} \in H^{2k}(BO; \Z_2)$.
 Given a fixed classifying map $f: X \to BO$ which represents a vector bundle
 $\xi$ over $X$, by a Wu structure on $\xi$ we mean a lifting 
 $\tilde{f}: X \to BO[v_{2k}]$ of the map $f$, $f= p\circ \tilde{f}$, where
 $p$ is the projection of the fibration. 
 The obstruction is obviously $v_{2k}(\xi)=f^*(v_{2k})$. If it vanishes then the set
 of all Wu structures on $\xi$, denoted by $Wu(\xi)$ or $Wu(f)$ obtains the natural structure of an 
 affine space over $H^{2k-1}(X;  \Z_2)$.  

\begin{definition}[\cite{BM} \cite{ML}]
A Wu structure on a space $M$ is a lifting of the classifying space map
$f: M \to BO$ to the connected cover $BO[v_i]$ obtained from $BO$ by killing
the class $v_i$. We have the following diagram 
\(
\xymatrix{
&&
BO[v_i]
\ar[d]^\pi
\\
M 
\ar@{-->}[urr]^{\hat{f}}
\ar[rr]^f
&&
BO
}\;.
\)
\end{definition}

The space $BO[v_i]$ is a principal fibration over $BO$
with fiber 
the Eilenberg-MacLane space $K(\Z_2, i-1)$
\(
\xymatrix{
K(\Z_2, i-1) 
\ar[rr]^=
\ar[d]
&& 
K(\Z_2, i-1)
\ar[d]
\\
BO[v_i] 
\ar[d]^\pi
\ar[rr]
&&
EK(\Z_2, i)
\ar[d]
\\
BO 
\ar[rr]^k
&&
K(\Z_2, i)
}\;,
\)
where the $k$-invariant of the fibration is an element $v_i$ in the 
cohomology $H^i(BO; \Z_2)$ defined by the $i$th Wu class  
$v_i$ of the universal bundle over BO.

\vspace{3mm}
Wu structures can be induced from other structures; for instance

\vspace{2mm}
\noindent {\bf (1)} {\it Spin structure}: A Spin structure leads to a Wu structure, 
because of the existence of the map $B{\rm Spin} \to BO[v_k]$. In fact, 
this holds for
many structures, including a framing and all connected covers $\cF$ of the orthogonal 
group such as String structure. This is captured by the commutative diagram 
\(
\xymatrix{
B \cF 
\ar[dr]
\ar[rr]^q
&&
BSO[v_k]
\ar[dl]^\pi
\\
&BSO&
}\;.
\label{eq BF}
\)
%\vspace{2mm}
\noindent {\bf (2)} {\it Dimension}:
Because of the fact that Wu classes vanish in degree above half the dimension 
of the manifold, we can have a Wu structure in the right degrees simply by 
this dimension argument. For example, on 
$(8k+2)$-dimensional spaces, the Wu class $v_{4k+2}$ is always zero \cite{BM}.
This is useful in studying the M5-brane as well as type IIB string theory 
(cf. \cite{S5}). 

%%%%%%%%%%
\section{Wu structures as twisted structures}
%%%%%%%%%%
\label{sec t}

The idea of this section is that the Wu structures themselves already 
can be interpreted as 
twisted structures corresponding to the Stiefel-Whitney classes of the
same degree. This provides connections to various other  structures. 
We describe the main point as follows. 
Rewrite relation \eqref{eq wu} as
\(
v_i= w_i -( Sq^1 v_{i-1} + \cdots + Sq^i v_0):= w_j - \alpha_i \quad {\rm for~} i=2^j\;,
\)
so that the terms involving the lower degree Wu classes can be 
interpreted as a twist for the structure defined by $w_i=0$. 
In low degrees, we see that this corresponds to twisted
Spin structure (for $j=1$) and twisted Membrane structure 
(for $j=2$). With this point of view, our definition is

\begin{definition}
Let $(X, \alpha_i)$ be a compact topological space with a degree $i$
cocycle $\alpha_i: X \to K(\Z_2, i)$ with $i=2^j$. 
A {\it Wu structure} over $X$ 
is a quadruple $(M, \nu , \iota, \eta)$, where

\noindent (1) $M$ is a smooth compact oriented manifold together with 
a fixed classifying map of its stable normal bundle $\nu: M \to BO$;

\noindent (2) $\iota: M \to X$ is a continuous map;

\noindent (3) $\eta$ is an $\alpha$-twisted structure on $M$ 
defined by $w_i(M) + \iota^* \alpha_i=0$, that is, a homotopy commutative diagram 

\(
    \raisebox{20pt}{
    \xymatrix{
       M
       \ar[rr]^\nu_>{\ }="s"
       \ar[d]_\iota
       &&
       B{\rm O}
       \ar[d]^{w_i}
       \\
       X
       \ar[rr]_{\alpha_i}^<{\ }="t"
       &&
       K(\Z_2,i)
       \ar@{=>}^\eta "s"; "t"
    }
    }
    \label{a t M2}
    \,,
\)
where $\eta$ is a homotopy between
$w_i \circ \nu$ and $\alpha \circ \iota$. 
\end{definition}

\paragraph{Remarks.} 
{\bf 1.} The above definition is inspired by definitions of other related structures in
\cite{Wa} \cite{SSS3} \cite{S4}.

\noindent {\bf 2.} In the above definition, we can replace $K(\Z_2, i)$ by
the product $K(\Z_2, i_1) \times K(\Z_2, i_2) \times \cdots \times 
K(\Z_2, i_r)$, where $i_1 + \cdots + i_r=i$. This product 
maps to $K(\Z_2, i)$.

\noindent {\bf 3.} Two Wu structures in this sense will be equivalent 
on $M$ if there is a homotopy between the corresponding 
homotopies $\eta$ and $\eta'$.

\vspace{3mm}
We record the idea above in the following

\begin{proposition}
A Wu structure in degree $2^i$ is a twisted structure for the structure defined by the
Stiefel-Whitney class in the same degree.
\end{proposition}

Our main example will be a Pin${}^-$ structure (see \cite{KT}), since
a $v_2$-structure is a Pin${}^{-}$ structure. The second Wu class $v_2$ is equal to 
the combination of Stiefel-Whitney classes $w_2 + w_1^2$.
If a  given manifold  $M$ is not Spin then $w_2(M)\neq 0$.
Since this takes values in $\Z_2$, the only nonzero value is 1.
This means that one can add to $w_2(M)$ a $\Z_2$-class 
$\alpha_1$, in this case equal to $w_1(M)^2$,
 such that $v_2= w_2(M) + \alpha_1=0 \in H^2(M; \Z_2)$.

\paragraph{Pin${}^-$ structures.}
A Pin${}^-$ structure on $M$ is equivalent to a Spin structure on 
$TM \oplus \det (TM)$.
The obstruction for existence of a Pin${}^-$ structure on $M$ is the 
characteristic class $w_2(M) + w_1(M)^2$. If $M$ admits a Pin${}^-$
structure, then the set of such structures ${\rm Pin}^-(TM)$ is acted upon freely 
and transitively by $H^1(M; \Z_2)$. 
In more detail, consider the two nontrivial central extensions 
$p_\pm: {\rm Pin}(n)^\pm \to O(n)$
of the orthogonal group $O(n)$ by $\Z_2$. A ${\rm Pin}^\pm$ structure on 
a vector bundle $E$ is a lifting of the structure group from $O(n)$ to its
double cover ${\rm Pin}(n)^\pm$.    
The short exact sequence
$
1 \to \Z_2 \to {\rm Pin}^\pm (n) 
\buildrel{p^\pm}\over{\longrightarrow}
O(n) 
\to 1
$
gives rise to an exact sequence 
\(
H^0(X, O(n)) 
\buildrel{\delta^0}\over{\to}
H^1(X; \Z_2)
\to
H^1(X, {\rm Pin}^\pm(n))
\buildrel{(p^\pm)^*}\over{\longrightarrow}
H^1(X, O(n)) 
\buildrel{\delta}\over{\to} 
H^2(X; \Z_2)\;.
\)
Applying the classifying functor $B$ gives
$B{\rm Pin}^\pm (n) 
\buildrel{Bp^\pm}\over{\longrightarrow}
BO(n)
\buildrel{\omega}\over{\to}
K(\Z_2, 2)$. Now a classifying map $f_E: X \to BO(n)$ of 
an $O(n)$ bundle $E$ has  a lift to $B{\rm Pin}^\pm (n)$
if and only if $\omega \circ f_E$ is homotopic to zero. Since 
$[X, K(\Z, 2)]\cong H^2(X; \Z_2)$, this is true when the 
the generators of $H^2(BO(n); \Z_2)$ pull back to zero in 
$H^2(X; \Z_2)$. These pullback classes are $w_2(E)$ and 
$w_2(E) + w_1(E)^2$.  The first corresponds to Pin${}^+$ 
structures and the second to Pin${}^-$ structures. See 
\cite{KT} for more details.

\vspace{3mm}
Let $0 \to E_1 \to E_2 \to E_3 \to 0$ be a short exact sequence of real vector
bundles. Let $\{i, j, k\}$ be a permutation of $\{1, 2, 3\}$. 
%\begin{enumerate}
%\item If $E_i$ and $E_j$ are Spin, $E_k$ has a natural Spin structure.
%\item If $E_i$ and $E_j$ are Spin${}^c$, $E_k$ has a natural Spin${}^c$ structure.
%\item If $E_i$ is Spin${}^c$ and $E_j$ is Pin${}^c$, $E_k$ has a natural 
%Pin${}^c$ structure.
%\item If $E_i$ and $E_j$ are Pin${}^c$, and $E_k$ is orientable, $E_k$ 
%has a natural Spin${}^c$ structure.
If $E_i$ is Spin and $E_j$ is Pin${}^\epsilon$, 
$E_k$ has a natural Pin${}^{-\epsilon}$ structure.
On the other hand, if $E_i$ is Pin${}^\epsilon$, $E_j$ is Pin${}^{-\epsilon}$,
and $E_k$ is orientable, 
$E_k$ has a natural Spin structure.

\paragraph{Example 4.}  Let $L$ be the real classifying line bundle over the real
projective space $\R P^m$ and let $x= w_1(L)$ be the generator of 
$H^1(\R P^m; \Z_2)=\Z_2$. Then 
$T (\R P^m) \oplus 1 = (m+1) \cdot L$, $w_1(\R P^m)= (m+1) \cdot x$ and 
$w_2(\R P^m)=\frac{1}{2} m (m+1)\cdot x^2$. Then
%$\R P^{4k+1}$ is Spin${}^c$.
$\R P^{4k+2}$ and $\R P^{4k+3}$ are both Pin${}^-$.
% and Pin${}^c$.
%$\RP^{4k+3}$ is Spin, Pin${}^\epsilon$, Spin${}^c$, and Pin${}^c$.
In fact, each 
%of $\R P^{4k+2}$ and  $\R P^{4k+3}$ 
admits two Pin${}^-$ structures. However, 
$\RP^{4k+3}$ is also Spin but $\R P^{4k+2}$ is not, since
even-dimensional real projective spaces are not orientable.

\vspace{3mm}
We have seen in section \ref{sec wu st} that Wu structures can 
result from Spin and other structures. Here we illustrate the point for
the Pin${}^-$ case. This will require replacing  BSO by BO
in diagram \eqref{eq BF}. 

\paragraph{Wu structures from Pin${}^-$ structures.}
Using the Adem relation $Sq^{4k+2}= Sq^2(Sq^{4k} + Sq^{4k-1} Sq^1)$, 
 in $H^*(BO; \Z_2)$ the following equality holds \cite{ML}
 \(
 v_{4k+2}= \sum_{i=1}^r a_i \cdot Sq^{I_i} v_2
\label{eq vv2}
 \)
 for certain classes $a_i \in H^*(BO; \Z_2)$ and certain Steenrod
 operations $Sq^{I_i} \in {\cal A}(2)$, the mod 2 Steenrod algebra, and $r \geq 1$. 
This gives a universal Wu structure for Pin${}^-$ bundles, so there exist
Wu structures on the universal vector bundle over $B{\rm Pin}^-$.
As described in \cite{Fin}, one of them can be fixed by taking a map
$
q: B{\rm Pin}^- \to BO[v_{4k+2}]
$
that completes the commutative diagram
\(
\xymatrix{
B{\rm Pin}^- 
\ar[rr]^q 
\ar[dr]
&&
BO[v_{4k+2}]
\ar[dl]^\pi
\\
&
BO
&
}
\)
with the natural projection $B{\rm Pin}^- \to BO$. 
Now, given a vector bundle $\xi$ over $X$ and its classifying map $f: X \to BO$, we have
the affine map 
$
q_f: {\rm Pin}^-(\xi) \to Wu(\xi)
$
which sends each lifting $\tilde{f}: X \to B{\rm Pin}^-$ to $q \circ \tilde{f}$. 
Here ${\rm Pin}^-(\xi)$ and ${\rm Wu}(\xi)$ are, respectively,
 the set of Pin${}^-$ structures
and Wu structures on the bundle $\xi$.

\paragraph{Special cases.}
When the composite Stiefel-Whitney classes are zero,  
the Wu class coincides with the Stiefel-Whitney class in that
degree -- at least for degrees that are powers of 2. 
In low degrees, Wu structures reduce to 

\vspace{2mm}
\noindent {\bf 1.} {\it Spin structure}: Here $v_2= w_2$ when $w_1=0$.

%\vspace{2mm}
\noindent {\bf 2.} {\it Membrane structure}: We have $v_4=w_4$ when $w_1=w_2=0$. 
This corresponds to the 
first Spin characteristic class $Q_1=\frac{p_1}{2}$, which is half the first Pontrjagin class $p_1$,
 being divisible by 2 \cite{S4}.

%\vspace{2mm}
\noindent {\bf 3.}  {\it Dual membrane structure}: 
In this case, $v_8=w_8$, when $w_1=w_2=w_4=0$. 
This corresponds to the second Spin 
characteristic class $Q_2=\frac{p_2}{2}$, which is half the second Pontrjagin class $p_2$,
 being divisible by 2
\(
Q_2= w_8 \mod 2\;.
\)
Working 2-locally, the vanishing
of this class also means we have a Fivebrane structure \cite{SSS2}.

%\begin{proposition} [Properties of Wu structures as twisted structures]
 %\end{proposition}

%

%%%%%%%%%%%
\section{Twisted Wu structures}
%%%%%%%%%%%
\label{sec tw}

A twisted Wu structure will be essentially 
defined by the condition $v + \alpha=0$. More specifically, 
a twisted Wu$(i)$ structure is defined by the condition $v_i + \alpha_i=0$, where
$\alpha_i \in H^i\Z_2$. 
Thus,
a twisted Wu($i$) structure on a manifold $X$ is described by the diagram 
\(
    \xymatrix{
       X
       \ar[rr]^f_>>>{\ }="s"
       \ar[drr]_{\alpha_i}^{\ }="t"
       &&
       B \mathrm{O}[v_j]
       \ar[d]^{v_i}
       \\
       &&
       K(\mathbb{Z}_2, i)~~,
       \ar@{=>}^\eta "s"; "t"
    }
  \)
  where $\eta$ is a homotopy between the map representing
  the class $v_i$ and the cocycle $\alpha_i$.  
More generally, we can also define such a twisted notion of Wu structure on a brane $M$.

\begin{definition}
Let $(X, \alpha_i)$ be a compact topological space with a degree $i$
cocycle $\alpha: X \to K(\Z_2, i)$. 
A {\it twisted Wu structure} over $X$ 
is a quadruple $(M, \nu , \iota, \eta)$, where

\noindent (1) $M$ is a smooth compact oriented manifold together with 
a fixed classifying map of its stable normal bundle $\nu: M \to BO$;

\noindent (2) $\iota: M \to X$ is a continuous map;

\noindent (3) $\eta$ is an $\alpha$-twisted structure on $M$ 
defined by $v_i(M) + \iota^* \alpha_i=0$, that is, a homotopy commutative diagram  
with a map $\iota: M \to X$
\(
    \raisebox{20pt}{
    \xymatrix{
       M
       \ar[rr]^\nu_>{\ }="s"
       \ar[d]_\iota
       &&
       B{\rm O}[v_j]
       \ar[d]^{v_i}
       \\
       X
       \ar[rr]_{\alpha_i}^<{\ }="t"
       &&
       K(\Z_2,i)~~,
       \ar@{=>}^\eta "s"; "t"
    }
    }
    \label{def tw wu}
\)
where $\eta$ is a homotopy between
$v_i \circ \nu$ and $\alpha \circ \iota$. 
\label{def tw}
\end{definition}

\paragraph{Remarks.} 
{\bf 1.} We have used $B{\rm O}[v_j]$ to indicate that we might have
some structure arising from Wu-connected cover of $B$O and not 
just $B$O itself (although in applications, the latter is more
dominant).

\vspace{2mm}
\noindent {\bf 2.}
When all the composite Stiefel-Whitney classes are zero (or all
Wu classes $v_j$ for $j<i=2^k$ are zero)  this reduces to $w_i + \alpha_i=0$.
Then, for $i=2$ and  8, we have a twisted Spin structure \cite{Wa} and a twisted Membrane
structure \cite{S4}, given respectively by $w_2 + \alpha_2=0$ and $w_4 + \alpha_4=0$.

%\vspace{3mm}
%This reduces in the Spin case to the twisted
 %membrane structures, defined in \cite{S4}. 

\paragraph{Example 5. Twisted Pin${}^-$ structures.} Let us consider 
the degree two case. We know that
$v_2$ is the obstruction for a Pin${}^-$ structure. When a manifold 
does not admit a Pin${}^-$ structure, this means that 
$v_2 \neq 0$. Since the Wu class is valued in $\Z_2$, 
being nonzero means it is $1\in \Z_2$. This implies that adding a nonzero class
in $H^2(M; \Z_2)$ will lead to a zero class for the sum, because this will
be 2-torsion. Therefore, if a manifold does not admit a Pin${}^-$ structure then 
it certainly admits what we call a {\it twisted Pin${}^-$ structure} 
(cf. definition \ref{def pin -} below). 
Examples of this include the real projective spaces
 $\RP^{4k}$ and $\RP^{4k+1}$. 

\begin{definition}
A twisted Pin${}^-$ structure is a twisted Wu structure in the sense of definition 
\ref{def tw} for $i=2$.
\label{def pin -}
\end{definition}

\paragraph{Example 6. Type IIA superstring theory with a B-field on an orientifold.} 
An orientifold is roughly a smooth manifold together with an involution and is
encoded in a double cover $\pi: X_\omega \to X$ of orbifolds, where $\omega \in H^1(X; \Z_2)$
is the equivalence class of the double cover. A B-field in this context is a differential 
cohomology class taking values in a certain Postnikov truncation of connected 
real K-theory $ko$ \cite{DisFM}. For type IIA string theory, one of the main results of \cite{DisFM} 
is that the first and second Stiefel-Whitney classes of 10-dimensional orientifold
spacetime $X$ are $w_1(X)= \omega$ and $w_2(X)= \omega ^2 + a(B) \cup \omega$, where
$a(B) \in H^1(X; \Z_2)$ is a certain class which accounts for the presence of 
 the B-field. Now we see that if $\omega \neq 0$ and $a(B) \neq 0$, i.e. if we have
a nontrivial orientifold and B-field, then the two expressions can be combined to give
$w_2(X) + w_1(X)^2= a(B) \cup \omega$. We can write this as $v_2 + \alpha_2=0$, so that
this structure indeed corresponds to a twisted Wu(2)-structure; in fact it corresponds to
the special case of twisted Pin${}^-$ structure described above.

\begin{observation}
A twisted Spin structure in the sense of \cite{DisFM} is a twisted Wu structure
in the sense of this paper, with the twist provided by both the orientifold double cover and
the B-field.
\end{observation}

\paragraph{Example 7. Type IIB string theory and twisted differential cocycles.}
The self-dual field in type IIB string theory is delicate and requires
some machinery to describe. In \cite{BeM} this is given by a Chern-Simons 
functional. Such a description requires dealing with 
differential integral Wu classes $\check{\lambda}$ 
which are
elements of the category
$\check{\mathcal{H}}_v^6$ of $\nu$-twisted differential 6-cocycles. This is  
 a torsor for differential characters $\check{\mathcal{H}}^6$ and 
whose objects are differential cocycles such that $v_6=a(\check{\lambda})$ mod 2,
where $a(\check{\lambda})$ is  the characteristic class  of $\check{\lambda}$.
Instead of viewing this as a character twisted by the Wu structure, we 
will turn it around and view it as a Wu structure twisted by the character. 
The reason for this is that the Wu class defines a bundle(-like) structure,
which should be the `main' structure, while the character is a slight 
modification. This is analogous to treating the $\frac{1}{4}p_1$-shifted
differential characters of \cite{DFM} as twisted String structure in \cite{SSS3}. 

\begin{observation}
A $v$-twisted differential character in the sense of \cite{BeM} is 
an instance of a twisted Wu structure in the sense of this paper.
\end{observation}

%%%%%%%%%%%%%%%
\section{Wu${}^c$ structures}
%%%%%%%%%%%%
\label{sec w c}

In this section we consider structures coming from Wu classes on which the 
action of the Bockstein vanishes. This Bockstein operation 
$\beta:H^i(X; \Z_2) \to H^{i+1}(X; \Z)$ 
is associated with the exact 
sequence of coefficients $0 \to \Z \buildrel{2}\over{\to} \Z \to \Z_2 \to 0$.
We define a Wu${}^c$ structure as an integral lift of a
 Wu structure. Therefore, a Wu${}^c$ class is defined via the
 Bockstein 
 on the corresponding Wu class, that is

% \paragraph{Defining Wu${}^c$ structures.}

 \begin{definition}
{(i).} A Wu${}^c$ characteristic class is given by 
$ V_{i+1}:= \beta v_i\in H^{i+1}(BO; \Z)$. This is the obstruction to 
having a Wu${}^c$ structure. 

\noindent { (ii).} A Wu${}(i)^c$ structure on a manifold $M$ is defined by 
$ V_{i+1}(M):= \beta v_i(M)=0\in H^{i+1}(M; \Z)$, with $v_i(M)=f^*v_i$
a pullback via the classifying map of the universal Wu class. 
That is, $v_i$ is the modulo 2 reduction of an integral class.
\label{def wu c}
 \end{definition}
 
\paragraph{Remarks.}
 {\bf 1.} The Wu${}^c$ characteristic class coincides with the integral Stiefel-Whitney class
 when the Wu class is indecomposable. 
% Note that this does not 
% occur in every dimension. For example, the expression for $v_6$ 
% does not involve $w_6$, so that in this case the Wu${}^c$ structure 
% does not reduce to the String${}^{K(\Z, 3)}$ structure define by the
% condition $W_7=0$. 
% 
%\vspace{2mm}

\vspace{2mm}
\noindent {\bf 2.} As in the case for integral Stiefel-Whitney classes, the
 Wu${}^c$ classes are also mostly relevant in the odd-dimensional
 case, i.e. for $j$ odd in $V_{j}$.
 However, for $\beta_2$ the Bockstein homomorphism coming from the
 exact coefficient sequence $0 \to \Z_4 \to \Z_4 \to \Z_2 \to 0$ instead 
 of the above sequence, the condition $w_1(M)^2=0$ is equivalent to 
 $\beta_2(w_1(M)=0$ and has application to manifolds of dimension 
 $4k+1$ \cite{DL}.

%Let us first concentrate on the oriented case.

\paragraph{Special cases.} We consider specializations of the Wu${}^c$
structure depending on the dimension. We will need to understand the action of the
Bockstein $\beta$ acting on products of Stiefel-Whitney classes. 
Let us start with the case when this product is a square.
Note that $\beta$ is an integral lift of $Sq^1$, which is a derivation and hence 
$Sq^1(x^2)=Sq^1(x) x + x Sq^1(x)=2x Sq^1(x)=0$. This means that
$\beta (x^2)$ must be a class that maps to 0 when reduced modulo 2. 
From the sequence 
$\Z \buildrel{\times 2}\over\longrightarrow \Z \buildrel{\rho_2}\over\longrightarrow \Z_2 
\buildrel{\beta}\over\longrightarrow \Z$ we see that 
if $\beta (x^2)=0$ mod 2, it must be twice a class, i.e. 
$\beta (x^2)=2y$ for some integral class $y$.
Furthermore, $y$ cannot be $\beta (z)$ for any mod class $z$, except 0, since
$2 \beta (z)=0$. Now since our classes are Stiefel-Whitney classes, 
we see that $\beta (w_2^2)=0$ since $w_2^2=\rho_2(p_1)$ and that
$\beta (w_4^2)=0$ since $w_4^2=\rho_2(p_2)$. Therefore, squares 
of Stiefel-Whitney classes vanish in our range of dimensions in the 
oriented cas. 
 
\vspace{2mm}
\noindent {\bf 1.} A Wu${}(2)^c$ structure is defined when  
$V_3=\beta v_2= \beta(w_2 + w_1^2)=0$, that is by 
$W_3 + \beta (w_1^2)=0$. 
If we impose orientation then this immediately gives a Spin${}^c$ structure. 
However, in full generality, the Wu$(2)^c$ structure can be interpreted as a 
Pin${}^c$ structure (see below).

\vspace{2mm}
\noindent {\bf 2.} A Wu$(4)^c$ structure is defined, in the oriented case, 
when $V_5=\beta v_4=\beta (w_4 + w_2^2)=0$. When we have the Spin condition
in addition, this reduces to the condition $\beta w_4=0$, which is the obstruction 
to a Membrane${}^c$ structure, introduced in \cite{S4}. Without the Spin 
requirement, a Wu$(4)^c$ structure might seem be a a twisted Membrane${}^c$
structure (defined in \cite{S4}) with the twist given by the integral class $\beta (w_2^2)$. 
However, as we saw above, $w_2^2$ admits an integral lift, and hence the
would be twist $\beta (w_2^2)$ is zero. 
We can consider this in more generality: the class $w_4 + w_2^2$ is in fact 
the mod 2 reduction of the class which defines a String${}^c$ structure
\cite{S3} (see below). An application for complex manifolds \cite{CSa}  gives 
$c_2 - c_1^2$ mod 2 $=w_4 + w_2^2 \in H^4(X; \Z_2)$.

\vspace{2mm}
\noindent {\bf 3.} A Wu$(6)^c$ structure is defined, in the oriented case, by imposing
$V_7=\beta v_6=\beta (w_2 w_4)=0$. 
Obviously, when we have either a Spin structure or a Membrane structure, then 
this condition is satisfied. In general, 
the Wu formula gives $w_6= Sq^2 w_4 + w_2 w_4$. At the level of 
Chern classes, this says that $\rho_2c_3=Sq^2(\rho_2c_2) + \rho_2(c_1c_2)$ mod 2.

\vspace{2mm}
We now summarize the main points above in the following

\begin{proposition}[Properties of Wu${}^c$ structures].

\noindent {\bf 1.} A $Wu(4)^c$ structure is the same as a  Membrane${}^c$ structure.

\noindent {\bf 2. (i)} 
A Spin structure implies a $Wu(6)^c$ structure.

{\bf (ii)} A Membrane structure implies a $Wu(6)^c$ structure.

\noindent {\bf 3.} The obstructions to Wu${}^c$ structures are additive for bundles, i.e.
$
(\beta v)(E \oplus F)
= \beta v(E) + \beta v(F)
$.
\end{proposition}

%Statements {\bf (ii)} and {\bf (iii)} in ${\bf 2}$ follow from the fact that a Spin structure implies a 
%Spin${}^c$ structure (or a Pin${}^+$ structure implies a Pin${}^c$ structure), and that
%a Membrane structure implies a Membrane${}^c$ structure. 
%The statement in ${\bf 3}$ follows from the fact that the Bockstein is a linear operation.

\paragraph{Application. Pin${}^c$ structure.}
Let $\rho_2: H^2(M; \Z) \to H^2(M; \Z_2)$ be the mod 2 reduction. 
%\begin{enumerate}
%\item $E$ is Spin $\leftrightarrow$ $w_1(E)=w_2(E)=0$.
%\item $E$ is Spin${}^c$ $\leftrightarrow$ $w_1(E)=0$ and $w_2(E) \in {\rm image}(\rho)$.
%\item $E$ is Pin${}^+$ $\leftrightarrow$ $w_2(E)=0$.
%\item $E$ is Pin${}^-$ $\leftrightarrow$ $w_2(E) + w_1(E)^2=0$.
 By definition, a bundle $E$ is Pin${}^c$ if and only if $w_2(E) \in {\rm image}(\rho_2)$, 
 that is $W_3(E)=\beta w_2(E)=0$ \cite{KT}.  
%\end{enumerate}
%Remark: 
As $w_1(E)^2= \rho_2 (c_1(E \otimes \C))$, 
$w_2(E) \in {\rm image}(\rho)$ if and only if $w_2(E)  + w_1(E)^2 \in {\rm image}(\rho_2)$
Note that if we have a sequence of bundles 
$0 \to E_1 \to E_2 \to E_3 \to 0$ with
$E_i$ Spin${}^c$ and $E_j$ Pin${}^c$, for $\{i, j, k\}$ a permutation of 
$\{1, 2, 3\}$, then $E_k$ has a natural 
Pin${}^c$ structure.
On the other hand, if $E_i$ and $E_j$ are Pin${}^c$, and $E_k$ is orientable, then 
$E_k$ has a natural Spin${}^c$ structure.

%\attn{Need similalry for Lusztig}

%$\beta v_2= W_3 + \beta (w_1) \cup w_1 - w_1 \cup \beta (w_1)$.

\paragraph{Example 8.}
The real projective spaces
$\R P^{4k}$,
 $\R P^{4k+2}$, and $\R P^{4k+3}$ are all
Pin${}^c$. However, the latter is also Spin, unlike the former two,
since real projective spaces of even dimension are non-orientable. 
 
\paragraph{Example 9. Wu${}^c$ structure in M-theory.}
Consider M-theory on an 11-dimensional Spin manifold $Y^{11}$
with a C-field, as described in \cite{DFM}. From the inclusion 
$H^4(Y^{11}; \Z_2) \hookrightarrow H^4(Y^{11}, \R/\Z) 
\hookrightarrow \check{H}^4(Y^{11})$, 
the cohomology class $w_4(Y^{11}) \in H^4(Y^{11}; \Z_2)$ 
defines a differential cohomology class $\check{w}_4$. 
The characteristic class of this flat character is the integral class
$W_5(Y^{11})= \beta w_4(Y^{11})$. 
This class is interpreted \cite{DFM} as 
 the background magnetic charge induced by the topology of $Y^{11}$
 and should vanish to be able to formulate the C-field. 
 On Spin manifolds, $W_5(Y^{11})=0$ since the class $\lambda$ is the
 integral lift of $w_4(Y^{11})$. Now, on a Spin manifold we have the fourth 
 Wu class $v_4=w_4$, so that we have as condition the vanishing of the
 integral lift of the Wu class $v_4$, i.e. 
 \(
 V_5(Y^{11})=\beta v_4(Y^{11})=0\;.
 \)
 Therefore, the C-field in M-theory leads to a Wu$(4)^c$ structure. 
 
 \vspace{3mm}
For manifolds of dimension $4k$, the characteristic elements for the intersection
pairing in the middle dimension are the integer lifts $\lambda$ of the
Wu-class $v_{2k}$.
 
\paragraph{Example 10. Wu${}^c$ structure for the M5-brane.} 
The study of the partition function of the M5-brane in \cite{W-5} \cite{HS} 
lifted to eight dimensions requires a mod 2 middle cohomology class
to lift to an integral class. 
If the M5-brane worldvolume if not Spin
(this happens often) then instead of $\beta w_4=0$ we will have
$\beta v_4=0$, that is $W_5 - \beta(w_2^2)=0$. This defines a Wu$(4)^c$ 
structure. We can also interpret this
as a twisted Membrane${}^c$ structure.

%%%%%%%%%%%
%\section{Integer lifts of Wu classes}
%%%%%%%%%%

\paragraph{Mod 2 reduction and Wu${}^c$ structure.}
Since the Steenrod square, the mod 2 reduction, and the 
Bockstein are related as $Sq^1= \rho_2 \beta$, we have 
$Sq^1 v_i=\rho_2(\beta v_i)$, so we have that a Wu${}^c$
structure implies that $Sq^1v_i=0$ in that degree. This latter
condition appears naturally when considering Wu classes
of Spin bundles (see Appendix in \cite{HS}): 
$Sq^1 v_{4k}(E)=0$ for a Spin bundle $E$. Conversely, 
this condition, which is always satisfied in the Spin case, 
implies that the mod 2 reduction of the Wu${}^c$ class
in degree $4k+1$ is zero. For example,
let $E$ be a real vector bundle over a space $X$  and 
consider the degree four Wu class $v_4$. Applying 
$Sq^1$ gives $Sq^1 v_4 U= Sq^1 \chi (Sq^4) U$, which 
by property of $\chi$ gives $\chi (Sq^4 Sq^1) U$. Now using the
Adem relation $Sq^2 Sq^3= Sq^4 Sq^1$, this gives
$\chi (Sq^2 Sq^3)U$. Applying the definition of $\chi$ leads to 
$\chi (Sq^3) Sq^2 U$. Since the bundle is assumed to be Spin
we have $Sq^2 U=w_2 U=0$, so that $Sq^1 v_4=0$.

 \vspace{3mm}
 We now consider the integral lifts of the Wu classes.
 We seek to characterize classes $x$ such that $\rho_2(x)=v$, according 
 to degrees.  
The discussion will proceed  according to whether
 the degrees are of the form $4k$ or $4k+2$. We start with the first case
 and illustrate for degree 4 and degree 8, the degrees which seem most 
 relevant for applications.

%
%Hopkins and Singer describe a family of integer cohomology classes
%$v_k^{\rm Spin}(E)$ for Spin bundles $E$, whose mod 2 reductions are
%the Wu classes of the underlying vector bundle. In degree 4, 
%$v_4^{\rm Spin}=-\frac{p_1}{2}$, rationally. 

\paragraph{String${}^c$ structures and integral lifts of the Wu class $v_4$.}
A String${}^c$ structure is defined by obstruction \cite{CHZ}
\(
Q_1 + c^2=0 \in H^4(M;  \Z)\;,
\)
where $c$ is the first Chern class of the complex line bundle 
which defines the Spin${}^c$ structure. On the other hand, in the oriented case
the Wu class $v_4$ is given in terms of the Stiefel-Whitney classes
as $v_4= w_4 + w_2^2$. Since $w_4$ is the mod 2 reduction of the 
first Spin class $Q_1=\frac{1}{2}p_1$ and $w_2$ is the mod 2 reduction
of the first Chern class $c$, we have that $w_4 + w_2^2$ is the 
mod 2 reduction of the   $Q_1 + c^2$. 
We have used the formula $w_2(M) \cup c= c \cup c$ mod 2 in the case
when our manifold is oriented, with $c \in H^2(M; \Z)$. 
Therefore, we have found 
an integral lift of the Wu class in this case

\begin{proposition}
An integral lift of a Wu(4)-structure on oriented 4-manifolds is given by a String${}^c$ structure.  
\end{proposition}

\paragraph{Fivebrane${}^{K(\Z, 4)}$ structures and integral lifts of the Wu class $v_8$.}
Now we consider a Fivebrane${}^{K(\Z, 4)}$ structure, defined by the condition \cite{S4}
\(
Q_2 + e^2=0 \in H^8(M; \Z)\;,
\)
where $e$ is the degree four class of a $K(\Z, 3)$ bundle. 
In the Spin case we have $v_8= w_8 + w_4^2$,
since $w_4 \cup e = e \cup e$ mod 2. Therefore, 
similarly to the degree 4 case above, we have

\begin{proposition}
An integral lift of a Wu(8)-structure on Spin 8-manifolds 
is given by a Fivebrane${}^{K(\Z, 4)}$ structure.  
\end{proposition}

 \vspace{3mm}
 We now consider the Wu classes of degree $4k+2$.
 
 \paragraph{Squares of odd-dimensional Steifel-Whitney classes and torsion Pontrjagin classes.}
 The Wu classes in degrees $2j$ contain squares of classes of degree $j$, i.e. $(w_j)^2$. 
 For low degree cases, this can be seen explicitly from \eqref{eq list}.
 We already understand the structures implied by the indecomposable 
 Stiefel-Whitney classes (and hence of corresponding Wu classes), at least 
 in lower degrees. The most notable decomposable terms not involving $w_1$
 are the squares mentioned above. Hence, we would like to gain a better 
 understanding of such terms.  For $j$ odd, we do this using torsion Pontrjagin classes
 \cite{Th}. These classes are $\mathcal{P}_{4k+2}$ with an indexing such that
 $\mathcal{P}_{4i}=p_i$ coincide with the usual $i$th Pontrjagin classes.
 For the degrees $4k+2$ (hence corresponding to half-integer indexing had we kept
  the usual notation) are 2-torsion: $2 \mathcal{P}_{4k+2}=0 \in H^{4k+2}\Z$. 
 One way to define these classes for a vector bundle $E$ is via the 
action of the Bockstein and the Steenrod square on the Stiefel-Whitney classes in 
degree $2k+1$, that is 
$\mathcal{P}_{4k+2}(E)= \beta Sq^{2k} w_{2k+1}(E)$. 
 The mod 2 reduction $\rho_2: H^{4k+2}(X; \Z) \to H^{4k+2}(X; \Z_2)$ of these classes
 gives precisely the desired squares of Stiefel-Whitney classes 
 $\rho_2 \mathcal{P}_{4k+2}(E)=w_{2k+1}(E)^2$.   
 Therefore, in the situation where the 
 the Wu class is given by the squares, we can find an integral lift. 
 
 \begin{proposition}
 In the situation described above, the integral lifts of the Wu classes
 are the torsion Pontrjagin classes. 
 \end{proposition} 
  
 \paragraph{Example 11.} We illustrate the proposition in low degree examples. 
 
 \noindent {\bf 1.} {\it Degree two}: $v_2=w_2 + w_1^2$. If $w_2=0$, that is we 
 have a Pin${}^+$ structure, then a lift of the Wu(2) structure is given by 
 the torsion Pontrjagin class $\mathcal{P}_2$. We could also be in a situation 
 where $w_2=\rho_2(c_1)$, where $c_1$ is the first Chern class. 
  
 \vspace{2mm} 
\noindent {\bf 2.} {\it Degree six}: In the oriented case we have $v_6=w_2w_4 + w_3^2$.
If $w_4=0$, i.e. if we have a Membrane structure \cite{S4} then 
$v_6$ reduces to the square term $w_3^2$. Then we see that the integral 
lift of the Wu class in this case if the torsion Pontrjagin class $\mathcal{P}_6$.
Note that we cannot instead set $w_2=0$ as then $w_3$ would also be zero. 

\vspace{2mm}
\noindent {\bf 3.} {\it Degree ten}:  Here $v_{10}$ will involve $v_2$, due to 
the formula \eqref{eq vv2}. Hence we cannot possibly isolate a square term. 
 However, if such a term is present then it will be given by the 
 torsion Pontrjagin class $\mathcal{P}_{10}$. Note that higher degree 
 Stiefel-Whitney classes can be disposed of, for example when considering 
 an oriented 11-manifold $Y^{11}$ for which $w_{11}(Y^{11})=w_{10}(Y^{11})=
 w_9(Y^{11})=0$
 (by using \cite{Ma}). 
 This gets rid of any possible terms involving $w_{10}$ and $w_9$ in $v_{10}$.

%%%%%%%%%%
\section{Twisted Wu${}^c$ structures}
%%%%%%%%%
\label{sec tw c}

In this section we will take the Wu${}^c$ structures we defined 
in the previous section and give them a twist. 
That is, we will consider a slight relaxation of the Wu${}^c$ condition.
A twisted Wu$(i)^c$ structure will be defined by the condition $\beta v_i + \alpha_{i+1}=0$,
where $\alpha_{i+1}$ is a degree $i+1$ integral class. More precisely, we 
have

\begin{definition}
Let $(X, \alpha_{i+1})$ be a compact topological space with a degree $i+1$
integral cocycle $\alpha_{i+1}: X \to K(\Z, i+1)$. 
A {\it twisted Wu${}^c$ structure} over $X$ 
is a quadruple $(M, \nu , \iota, \eta)$, where

\noindent (1) $M$ is a smooth compact oriented manifold together with 
a fixed classifying map of its stable normal bundle $\nu: M \to BO$;

\noindent (2) $\iota: M \to X$ is a continuous map;

\noindent (3) $\eta$ is an $\alpha$-twisted structure on $M$ 
defined by $V_{i+1}(M) + \iota^* \alpha_{i+1}=0$, that is, a homotopy commutative diagram  
\(
    \raisebox{20pt}{
    \xymatrix{
       M
       \ar[rr]^\nu_>{\ }="s"
       \ar[d]_\iota
       &&
       B{\rm O}[v_j]
       \ar[d]^{V_{i+1}}
       \\
       X
       \ar[rr]_{\alpha_{i+1}}^<{\ }="t"
       &&
       K(\Z,i+1)~~,
       \ar@{=>}^\eta "s"; "t"
    }
    }
    \label{def tw wu c}
\)
where $\eta$ is a homotopy between
$w_i \circ \nu$ and $\alpha \circ \iota$. 
\label{def Vi}
\end{definition}

\paragraph{Remarks.}
{\bf 1.} The idea of the twisted Wu$(i)^c$ structure is that the Wu class might not
be the mod 2 reduction of an integral class exactly, but only so up to an 
(auxiliary) integral class. 

\vspace{2mm}
\noindent {\bf 2.} If the composite Stiefel-Whitney classes all vanish, then 
a twisted Wu$(i)^c$ structure reduces to a twisted Pin${}^c$ structure,
twisted Membrane${}^c$ structure, and twisted String${}^{(K(\Z, 3)}$
 structure for $i=2, 4$, and 6, respectively. The latter two are defined
 in \cite{S4}, and so we now give a definition of the first.

\begin{definition}
A twisted Pin${}^c$ structure is a structure given in definition \ref{def Vi}
for $i=2$. 
\end{definition}
When the twist is zero, this reduces to the usual Pin${}^c$ structure.

\vspace{3mm}
We now consider a sample property of the twisted Wu${}^c$ structures. 
The obstructions to twisted Wu${}^c$ structures are additive for bundles
\bea
(\beta v + \alpha)(E \oplus F)&=& \beta v(E \oplus F) + \alpha(E \oplus F)
\nonumber\\
&=& \beta v(E) + \beta v(F) + \alpha(E) + \alpha(F)\;,
\eea
provided that $\alpha$ is an additive class, as then we have
\(
\beta v(E \oplus F) = \beta v(E) + \beta v(F)\;.
\)

\begin{proposition}
Wu${}^c$ structures are additive. 
\end{proposition}

We will find that for our applications we need a relative version of 
twisted Wu${}^c$ structures. Relative versions of classical Wu classes
are defined in \cite{Ker}, on which we build our generalization.

\paragraph{Relative Wu classes.}
Consider a manifold $Z$ with boundary $Y$ and corresponding tangent bundles 
$\pi: TZ\to Z$ and $p: TY\to Y$. The relative second
Wu class $v_2(Z, Y)$ can be defined via the diagram
\(
\xymatrix{
H^{1}(TY; \Z_2)
\ar[r]^{\hspace{-3mm}\delta_T}
&
H^2(TZ, TY; \Z_2)
\\
H^{1}(Y; \Z_2)
\ar[u]^{\pi^*}
\ar[r]^{\hspace{-2mm}\delta}
&
H^2(Z, Y; \Z_2)
\ar[u]^\cong_{\pi^*}
}
\label{diag rel w}
\)
i.e. is the element $(\pi^*)^{-1} \circ \delta_T (\sigma)$,
where $\sigma$ is an element of the set of Pin${}^-$ 
structures on $TY$. 
If $j: (Z, \emptyset) \to (Z, Y)$ is the inclusion then 
$j^*v_2(Z, Y; \sigma)=v_2(Z)$ is the usual class for each
$\sigma$ in the set of Pin${}^-$ structures ${\rm Wu}_2(Y)$.
The following proposition is a straightforward extension
of a result in \cite{Bo} in the Spin case.

\begin{proposition}
The function $v_2(Z, Y): {\rm Wu}_2(Y) \to H^2(Z, Y; \Z_2)$ satisfies the 
following:

\noindent (i) image$(v_2(Z, Y))=(j^*)^{-1}(v_2(Z))$;

\noindent (ii) if $Z$ is 4-dimensional, 
a class $v \in H^2(Z, Y; \Z_2)$ lies in the image of 
$v_2(Z, Y)$ if and only if $\langle v, j_*(\xi)\rangle
\equiv \xi \cdot \xi$ (mod 2) for each $\xi \in H_2(Z)$;

\noindent (iii) if $x \in H^1(Y; \Z_2)$ and $\sigma \in {\rm Wu}_2(Y)$, then 
$$
v_2(Z, Y; x\cdot u)= v_2(Z, Y; \sigma) + \delta x.
$$
Thus $v_2(Z, Y)$ is injective;

\noindent (iv) if $Z$ is 4-dimensional, then $v_2(Z, Y; \sigma)=0$ if and only if 
the intersection pairing on the middle homology of $Z$ is even and 
$\sigma=\sigma_Z$. 

\end{proposition}

\vspace{3mm}
The proof of (ii) follows from (i) and an application of the Wu formula. 
The latter also establishes the rest of the statements.

\vspace{3mm}
Note that by Lefschetz duality $D: H^i(Z, Y; \Z_2) \to H_{n-i}(Z; \Z_2)$
one gets an absolute homology class which is dual to the relative
cohomology class; hence for some purposes it might be more convenient 
to work with homology. We use this in the propsitions below; see \cite{S5}
for more extensive applications.

\vspace{3mm}
The above discussion on relative Pin${}^-$ structures can be generalized
to higher Wu structures. Our examples will keep us grounded and consider
relatively low degree cases instead 
 of aiming for utmost generality.
We will need the following to consider the M5-brane case.

\paragraph{Relative Wu class on eight-manifolds.}
Consider an 8-manifold $Z^8$ with boundary $Y^7$ and tangent bundles 
$\pi: TZ^8\to Z^8$ and $p: TY^7\to Y^7$. The relative fourth
Wu class $v_4(Z^8, Y^7)$ can be defined via the diagram
\eqref{diag rel w}, with the obvious changes in the degrees of the
cohomology groups, This is  
the element $(\pi^*)^{-1} \circ \delta_T (\sigma)$,
where $\sigma$ is an element of ${\rm Wu}_4(Y^7)$, 
the set of ${\rm Wu}_4$ 
structures on $TY^7$. Similarly, we have

\begin{proposition}
The function $v_4(Z^8, Y^7): {\rm Wu}_4(Y^7) \to H^4(Z^4, Y^7; \Z_2)$ satisfies the 
following:

\noindent (i) image$(v_4(Z^8, Y^7))=(j^*)^{-1}(v_4(Z^8))$;

\noindent (ii) a class $v \in H^4(Z^8, Y^7; \Z_2)$ lies in the image of 
$v_4(Z^8, Y^7)$ if and only if $\langle v, j_*(\xi)\rangle
\equiv \xi \cdot \xi$ (mod 2) for each $\xi \in H_4(Z^8)$;

\noindent (iii) $v_4(Z^8, Y^7; \sigma)=0$ if and only if 
the intersection pairing on the middle homology of $Z^8$ is even and 
$\sigma=\sigma_Z$. 
\end{proposition}

\paragraph{Example 12. Type IIB string theory.} 
Similar results holds for the 6th Wu class on 12-manifolds with
boundary, relevant to type IIB string theory.

\vspace{3mm}
We now need a notion of relative ${\rm Wu}^c$-strucutre. 
A straightforward application of the general formulation in \cite{Ker}
to our ${\rm Wu}^c$ classes (cf. Definition \ref{def wu c})
gives the following defintion

\begin{definition}
 For a pair of spaces $(Z, Y)$, we define a relative 
${\rm Wu}^c$ structure by applying the Bockstein on the 
relative Wu classes, that is $V_{i+1}(Z, Y)= \beta v_i (Z, Y)$.
\end{definition}

Next, we add the twist to our relative ${\rm Wu}^c$ structures. 

\begin{definition}
 A relative twisted Wu${}^c$) structure on a pair $(Z, Y)$ 
is described by the diagram 
\(
    \xymatrix{
       (Z, Y)
       \ar[rr]^f_>>>{\ }="s"
       \ar[drr]_{\alpha_{i+1}}^{\ }="t"
       &&
       B \mathrm{O}[v_j]
       \ar[d]^{V_{i+1}}
       \\
       &&
       K(\mathbb{Z}, i+1)~~,
       \ar@{=>}^\eta "s"; "t"
    }
  \)
  where $\eta$ is a homotopy between the map representing
  the relative integral class $V_{i+1}(Z, Y)$ and the 
relative cocycle $\alpha_{i+1}$.  
\end{definition}

\paragraph{Remarks.}

{\bf 1.} As in previous defintions, such a twisted notion of Wu structure can be 
defined on a brane $M$ with the (by now) obvious changes. 

\vspace{2mm}
\noindent {\bf 2.} The condition for existence of a relative twisted
Wu${}^c$ structure of degree $i$ is $V_{i+1}(Z, Y) + \alpha_{i+1}=0\in H^{i+1}(Z, Y; \Z)$.

\vspace{3mm}
With the above defintions and results, we now present our main application of
relative twisted Wu${}^c$ structures.

\paragraph{Example 13.  The M5-brane.}
Consider the worldvolume of the M5-brane as a 6-manifold 
$M^6$. In the usual Chern-Simons construction, this is the base of 
a circle bundle $Y^7$, which is a boundary of an 8-manifold $Z^8$. 
The fields are considered to be cohomology classes 
of degree three (i.e. middle-degree) on $M^6$ which lift to degree
four cohomology classes on $Z^8$. In addition to the usual 
intersection pairing on $M^6$, we have a torsion pairing on $Y^7$.
Let $T^4(Y^7)$ denote the torsion subgroup of the integral cohomology
group $H^4(Y^7; \Z)$, and similarly let $T^4(Z^8)$ be the torsion 
subgroup of $H^4(Z^8; \Z)$. The torsion pairing is given by  the symmetric
bilinear nonsingular pairing 
\(
L: T^4(Y^7) \times T^4(Y^7) \to \Q/\Z\;.
\)
Let $i: T^4(Z^8) \to T^4(Y^7)$ be the inclusion. This is the adjoint with
respect to the pairing $L$ of the map $\delta: T^4(Y^7) \to T^5(Z^8, Y^7)$,
that is $L(i(x), y)=\langle x, \delta y \rangle$, where $x\in T^4(Z^8)$ and 
$y \in T^4(Y^7)$. 

\vspace{3mm}
A lifting $\hat{v} \in H^4(Z^8, Y^7; \Z_2)$ of the Wu class $v_4 \in H^4(Z^8; \Z_2)$
is said to be compatible with the quadratic form $\psi: T^4 (Y^7) \to \Q/\Z$ if, for
all $x \in T^4(Y^7)$ such that $x=i^*(y)$, we have
\(
\psi (x)= \frac{1}{2} \langle y\cdot \hat{v}_4, [Z^8, Y^7] \rangle
-\frac{1}{2} \langle y \cdot (j^*)^{-1}y, [Z^8, Y^7]\rangle \in \Q/\Z\;.
\)
Let $j^*: H^{4}(Z^8, Y^7; \Z) \to H^4(Z^8; \Z)$ be the map induced from the map that
forgets the boundary. 
Given a quadratic function on $T^4(Y^7)$, there is a corresponding 
Wu class $\hat{v}_4 \in H^4(Z^8, Y^7; \Z_2)$. Now let $b_4\in T^4(Y^7)$ be 
class such that $\psi i (y)= L(b_4, i(y))$  for all $y \in T^4(Z^8)$. Then we have
\(
\beta \hat{v}_4 = \delta^* b_4\;,
\label{eq b4}
\)
where $\beta: H^4(Z^8, Y^7; \Z_2) \to H^5(Z^8, Y^7; \Z)$ is the integral
Bockstein and $\delta^*: H^4(Y^7; \Z) \to H^5(Z^8, Y^7; \Z)$ is the 
coboundary map of the  pair $(Z^8, Y^{7})$.

\vspace{3mm}
We now interpret expression \eqref{eq b4} as defining 
a twisted (relative) Wu$(4)^c$ structure.

\vspace{3mm}
There exists an integral class $v_4'\in H^4(Z^8; \Z)$ such that
the mod 2 reduction is the absolute (i.e. non-relative)
Wu class $\rho_2(v_4')= v_4 \in H^4(Z^8; \Z_2)$ and 
$i^*(v_4')= 2b_4 \in H^4(Y^7; \Z)$.

%\paragraph{Example. The M5-brane.}
%Consider the M5-brane worldvolume as a Spin manifold $X^{6}$, 
%considered as the base of a circle bundle $Y^{7}$,
%which is the boundary of an eight-manifold $Z^{8}$.
%Defining the partition function properly requires working 
%with a Wu class $v_4 \in H^4(Z^{8};\Z_2)$ and a 
%function $\psi: T^4(Y^{7}) \to \Q/\Z$, where $T^4 \subset H^4(Y^{7};\Z)$ 
%is the torsion subgroup. 
%Consider a lifting $\hat{v}_4 \in H^4(Z^{8}, Y^{7};\Z)$ of the Wu class $v_4$.
%From \cite{BM}, compatibility of this lifting with the function $\psi$ implies that
%there exists a class $b\in T^4(Y^{7})$ 
%such that $\beta \hat{v}_4=\delta^* b$, where $\beta: H^4(Z^{8}, Y^{7};\Z_2) 
%\to H^5(Z^{8}, Y^{7};\Z)$ is the integral Bockstein and 
%$\delta^*: H^4(Y^{7};\Z) \to H^5(Z^{8}, Y^{7};\Z)$ is the coboundary map of the 
%pair $(Z^{8},  Y^{7})$.
 
 \vspace{3mm}
Consider the Wu relation $Sq(v)=w$, where $Sq$ is the total Steenrod square operation, 
$v$ is the total Wu class and  $w$ is total Stiefel-Whitney class. The degree 4 component 
gives $v_4=w_4 + $ lower degree classes.  If we require these lower degrees to vanish, then
the Wu class and the Stiefel-Whitney class coincide in degree 4. Therefore, the above 
condition can be written as 
\(
\beta w_4= \delta^* b\;,
\)
which is of the form $W_5 - H_5=0$ (albeit in relative cohomology).

\paragraph{Example 14.  Type IIB string theory.} 
The structure of type IIB string theory in ten dimensions, on $M^{10}$, 
is in some ways very similar to that
of the M5-brane. In particular, here we use for fields cohomology
classes of degree five in ten dimensions which lift to cohomology
classes of degree six in twelve dimensions, on 
$Z^{12}$ with $\partial Z^{12}=Y^{11}$. In this case, the 
above construction yields the condition in degree seven
\(
\beta \hat{v}_6 = \delta^* b_6\;,
\label{eq b6}
\)
where $b_6 \in T^6(Y^11)$ be 
class such that $\psi i (y)= L(b_6, i(y))$  for all $y \in T^6(Z^{12})$.

\vspace{3mm}
We now interpret expression \eqref{eq b6} as defining 
a twisted (relative) Wu$(6)^c$ structure.

\vspace{3mm}
There exists an integral class $v_6'\in H^6(Z^{12}; \Z)$ such that
the mod 2 reduction is the absolute (i.e. non-relative)
Wu class $\rho_2(v_6')= v_6 \in H^6(Z^{12}; \Z_2)$ and 
$i^*(v_6')= 2b_6 \in H^6(Y^{11}; \Z)$.

\vspace{3mm}
We record the results of the above two examples in the following

\begin{proposition}
The lift of the Wu class $v_i$ in the sense of \cite{BM} gives rise to a twisted 
Membrane${}^c$ structure and a twisted String${}^{K(\Z,3)}$ 
structure on the 8-manifold and the 12-manifold for $i=4$ and 
$i=8$, respectively.
\end{proposition}

Example 14 is also discussed in \cite{S5} in relation to global 
anomalies in type IIB string theory.
% and in \cite{SW} in relation to 
%twisted spectra at chromatic level two. 

\vspace{0.5cm}
\noindent {\large \bf Acknowledgements}

\vspace{2mm}
The author would like to thank IHES, Bures-sur-Yvette, for 
kind hospitality and inspiring atmosphere during the writing 
of this paper. This work is supported by NSF Grant PHY-1102218.

%%%%%%%%%%%%%%%%%%%%%%%%%%%%%%%%%%%%%%%%%%%%%

\end{document}